\documentclass[10pt,letterpaper,twocolumn]{article} %% two column, final layout

\usepackage{ol2}
\usepackage[draft,implicit=false]{hyperref}
\usepackage{amsmath}

\usepackage{amsmath,amssymb,graphicx}

\def\note#1{} %use this to make comments disappear !

\def\medskip{}

\begin{document}

\twocolumn[ %% activate for two-column option

\title{Comb--assisted coherence transfer between laser fields}

\author{
Tommaso Sala$^1$, Samir Kassi$^{2,3}$, Johannes Burkart$^{2,3}$, Marco Marangoni$^1$, Daniele Romanini$^{2,3,*}$
}

\address{
$^1$Physics Department of Politecnico di Milano and IFN-CNR, Piazza Leonardo da Vinci 32, 20133 Milano, Italy\\
$^2$Univ. Grenoble Alpes, LIPhy, F-38000 Grenoble, France\\
$^3$CNRS, LIPhy, F-38000 Grenoble, France\\
$^*$Corresponding author: daniel.romanini@ujf-grenoble.fr
}

\begin{abstract}
Single mode laser fields oscillate at frequencies well outside the realm of electronics, but their phase/frequency fluctuations fall into the radio frequency domain, where direct manipulation is possible. Electro--optic devices have sufficient bandwidth for controlling and tailoring the dynamics of a laser field down to sub--nanosecond time scales. Thus, a laser field can be arbitrarily reshaped and in particular its phase/frequency fluctuations can be in principle removed. In practice, the time evolution of a reference laser field can be cloned to replace the fluctuations of another laser field, at a close-by frequency.
In fact, it is possible to exploit a partially stabilized optical comb to perform the cloning across a large frequency gap. We realize this long--haul phase transfer by using a fibered Mach--Zehnder single--sideband modulator driven by an appropriate mix of the beat notes of the master and the slave laser with the comb.
\end{abstract}


] %% activate for two-column option

Frequency and time metrology foundations rest today on optical frequency combs (OFC) working as frequency rulers across an extremely wide spectral domain. Applications range from referencing ultrastable continuous--wave (CW) laser sources\cite{Jones2000,Diddams2000a} to high precision spectroscopy\cite{Truong2013}. An OFC features a broad spectral envelope filled by narrow uniformly spaced modes with frequencies $f_n = f_0 + n f_{rep}$, where $n$ is an integer. The mode spacing $f_{rep}$ and the comb offset $f_0$ fall in the radio frequency (RF) domain and can be measured and controlled to high accuracy\cite{Jones2000}.

Mixing a CW laser with an OFC on a fast photodiode delivers a beat note of its emission line with the closest comb mode. Given preliminary approximate knowledge of the laser frequency (to better than $f_{rep}/2$), a highly refined and accurate frequency measurement is then obtained from a knowledge of $f_0$ and $f_{rep}$. An ultrastable comb can thus become a ruler for locking and narrowing a CW laser emitting anywhere inside its wide spectral envelope. However, one or even two ultrastable CW lasers are then needed to stabilize the comb\cite{Bartels2004}. In the end, such general schemes of comb--mediated coherence transfer rely on a few high speed phase--locked loops to achieve referencing of a CW slave laser to one or two high coherence CW master lasers across a wide spectrum and require full comb stabilization\cite{Ruehl2011}.

In order to completely transfer coherence, the bandwidth and the control dynamic range of a servo loop must be sufficient to handle the phase--frequency noise spectrum of the free running laser. The requirement on the dynamic range is even stronger in a noisy environment with severe $1/f$ noise, i.e. outside a metrology laboratory.
Laser setups involving high performance phase locking loops between several lasers are expensive and difficult to build and maintain, and demand intervention on the lasers to be stabilized by introducing fast frequency control means, e.g. electro-optic or acousto-optic modulators, which may have to be installed inside the laser cavity.

So-called feed--forward (FF) schemes constitute a simpler approach: Phase/frequency variations derived from a fraction of the laser radiation are subtracted from the main laser beam, or else from a phase sensitive RF signal generated from it. For instance, frequency mixing in the RF domain has been used to subtract from a beat signal the $f_0$ note produced by an $f-2f$ OFC beating system\cite{Telle2002}, avoiding stabilization of $f_0$. In the optical domain, an AOM acting as a frequency shifter allowed to subtract from an OFC beam its $f_0$ fluctuations\cite{Koke2010}, or to subtract the frequency fluctuations of a CW laser relative to an OFC\cite{Sala2012,Gatti2012}. In both applications the RF signal driving the AOM was the beat note itself. These optical FF approaches have a sub-MHz control bandwidth due to the acoustic wave propagation delay in the AOM.

We show here that coherence transfer between two CW lasers lying inside the wide frequency range of an OFC can be obtained using a only one external control device to apply a FF phase correction to the slave laser. This correction has a double effect of eliminating at the same time ``common--mode'' fluctuations of the comb at the two lasers frequencies, as well as fluctuations of the slave with respect to the master laser.
This simple FF scheme allows for larger control bandwidth and dynamic range than any previously demonstrated approach, due to a fibered EOM device operated as a single--sideband modulator\cite{Burkart2013}. It can be applied to any CW single mode laser and even used to copy complex field dynamics from the master to the slave laser. It does not suffer from environmental perturbations, i.e. it cannot ``unlock'' as it may occur to any feedback--based control loop under a violent perturbation. Thanks to these properties it allows virtually instantaneous transfer of a frequency tuning action from the master to the slave, contrary to lock loops which can tolerate frequency changes at a given maximum rate. In fact, phase lock servo systems developed in metrology laboratories to deliver high frequency stability are not intended for nor easily adapted to frequency agile applications. For this reason they are not easily seen in a molecular spectroscopy laboratory.
Besides, contrary a feedback loop, the proposed approach does not require acting on the slave laser, which can perturb its operation for instance by exacerbating laser amplitude noise due to laser phase--amplitude couplings or by increasing phase noise at frequencies just outside the loop bandwidth.
The principal drawback of our FF approach is the low power level available in phase modulation sidebands. This is not an issue for applications sensitive to source coherence, e.g. spectroscopy in a high finesse cavity, where the loss in total power is compensated by an increase of the power spectral density in the carrier (proportional to cavity injection efficiency).

\begin{figure}
\centerline{\rotatebox{0}{\scalebox{0.47}{\includegraphics{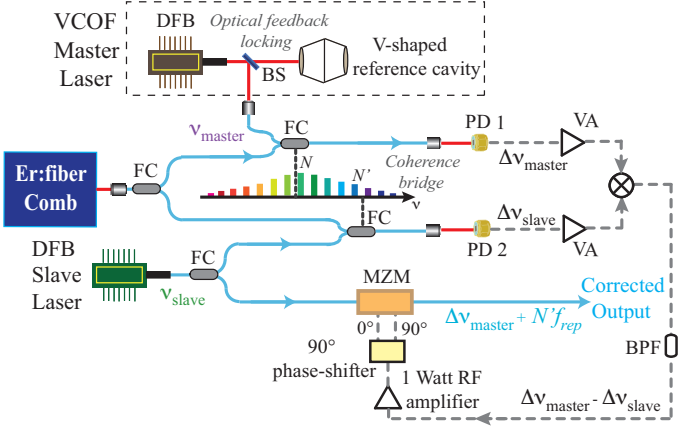}}}}
\caption{\label{fig:experimental} Experimental scheme. VCOF is the sub-kHz linewidth master laser, DFB the slave distributed--feedback diode laser, OFC the self-referenced optical frequency comb, MZM the Mach-Zehnder electro--optic single--sideband modulator, PD1 and PD2 are photodiodes collecting the beat notes of both lasers with the OFC, filtered by diffraction gratings (DG). FC are single mode fiber combiners/splitters, VA are RF voltage amplifiers, BPF is a bandpass RF filter.
}
\end{figure}

This scheme of coherence transfer has high potential for applications in frequency metrology, but also in high precision spectroscopy as it allows agile and exact frequency tuning.%not only by tuning the master laser, but also by simply adding a synthesized RF frequency shift to the FF correction signal.
Our goal is to improve the frequency accuracy of ultrasensitive absorption measurements using fibered DFB diode lasers coupled with cavity ring-down spectroscopy (CRDS). Over recent years, CRDS generated refined lists of thousands spectral lines of molecules of atmospheric or planetological interest. Nonetheless, while CRDS is linear over 5 decades of the absorption scale\cite{Burkart2014}, its frequency scale is limited by the $\sim20$\,MHz accuracy of the best commercial wavelength meter.
Towards this goal, we recently developed a source of high coherence and stability based on a DFB diode laser locked to an isolated high finesse V-shaped cavity by optical feedback (VCOF)\cite{Burkart2013}. Frequency tuning is provided by a fibered monolithic dual--parallel Mach--Zehnder electro--optic modulator (MZM) used as a single--sideband modulator with excellent ($\sim30$\,dB) suppression of the carrier and of other sidebands\cite{Burkart2013}.

%We introduce here a comb--assisted FF approach to ``clone'' the sub-kHz linewidth of such a stable source to any other DFB diode laser inside the comb spectrum, and provide accurate and wide frequency tuning (by RF synthesis) with efficient injection of a high finesse ring--down cavity.

We recently demonstrated coherence transfer or ``phase cloning'' between two CW lasers by MZM--based FF correction applied to a DFB diode laser\cite{Bukart2014a}, worse--case of a noisy laser. As the beat note between slave and master lasers gives their instantaneous frequency difference, this is subtracted from the slave laser radiation by applying the beat signal directly to the MZM single--sideband modulator, which works as a fast (GHz) frequency shifter: The corrected radiation is a clone of the master laser field.

That FF scheme can be extended by introducing an OFC as a coherent bridge between master and slave lasers, which scales it up from GHz to THz frequency gaps. For simplicity, let us assume we select with suitable RF band-pass filters the beat notes of the two lasers, each with the closest lower--frequency comb modes:  $\Delta\nu_m$ for the master laser and $\Delta\nu_s$ for the noisy slave laser. A mixer provides their difference, $\Delta\nu_m - \Delta\nu_s = \nu_m-(f_0+Nf_{rep})-[\nu_s-(f_0+N'f_{rep})] = \nu_m - \nu_s + (N'-N)f_{rep}$, which is amplified and applied to the MZM, traversed by a fraction of the slave laser radiation. In this way, after adjusting the MZM control bias voltages to produce only the upper sideband, the RF signal driving this sideband will subtract the slave laser frequency fluctuations relative to the comb modes and, at the same time, the common--mode $f_0$ frequency fluctuations of the comb modes relative to the master laser. In equations, the MZM output shall be at frequency $\nu_c = \nu_s + [\nu_m - \nu_s + (N'-N)f_{rep}] = \Delta\nu_m + N'f_{rep}$, thus being at the same comb offset as the master laser but relative to the comb mode $N'$. As the separation of slave from master is increased, the comb frequency fluctuations that are not common mode at the two laser frequencies, are expected to increase and to become a limiting factor for the coherence transfer. This effect is small and goes undetected in our proof-of-principle demonstration using a commercial partially stabilized OFC. It is found to be negligible in applications involving laser injection of a high finesse CRDS cavity, over a comb--wide frequency gap. Also, as with active servo loops, the bandwidth of this FF control is inversely proportional to time delays accumulated by both electrical and optical signals. However, the latter may be used to compensate the former provided a fiber patch of suitable length is added to the slave propagation path upstream the MZM correction unit\cite{Bukart2014a}: To first order this leads to a zero net delay. The bandpass is then limited by photodetectors and RF components used to drive the MZM.

Fig.\ref{fig:experimental} resumes our experimental implementation. Besides the VCOF source detailed elsewhere\cite{Burkart2013} and a standard fibered telecom DFB diode laser, a 100\,MHz self-referenced Erbium fiber comb (Toptica FFS model) is used, with free-running mode linewidth of about 10\,kHz over 1\,ms.
This OFC is partially stabilized in the sense that moderate bandwidth control loops are applied to obtain a long term stability of $f_0$ and $f_{rep}$: Only the low frequency drift and jitter of the comb modes are corrected while their short term linewidth maintains its free--running value. Care is taken to obtain beat signals with good S/N (in excess of 25\,dB) for both CW lasers. The $\Delta\nu_s$ beat is centered around 10-20\,MHz and low--pass filtered at 30\,MHz, while $\Delta\nu_m$ is kept around 70\,MHz at the center of a narrow bandpass filter. The sign of the beat notes is chosen in order to use the difference of these signals as the driving signal of the MZM (consistently with our equations above). RF amplification ($\sim$25-40\,dB) brings up both $\Delta\nu_m$ and $\Delta\nu_s$ signals to the range where the mixer operates with negligible noise addition.
A bandpass filter selects the difference of the beating notes after the mixer. A 1\,Watt RF amplifier boosts this signal up to a level adequate for the MZM to deliver 2-3\% of the incident optical power into the first sideband, and negligible power in higher order harmonics. We should note that together with the carrier, destructive interference in the dual--MZM eliminates all even order sidebands as well\cite{Bukart2014a}. The amplified signal is applied to the MZM via a 90$^\circ$ RF splitter operating in the range from 55 to 90\,MHz.

\begin{figure}
%\centerline{
\rotatebox{0}{\hspace{-4mm}\scalebox{0.37}{\includegraphics{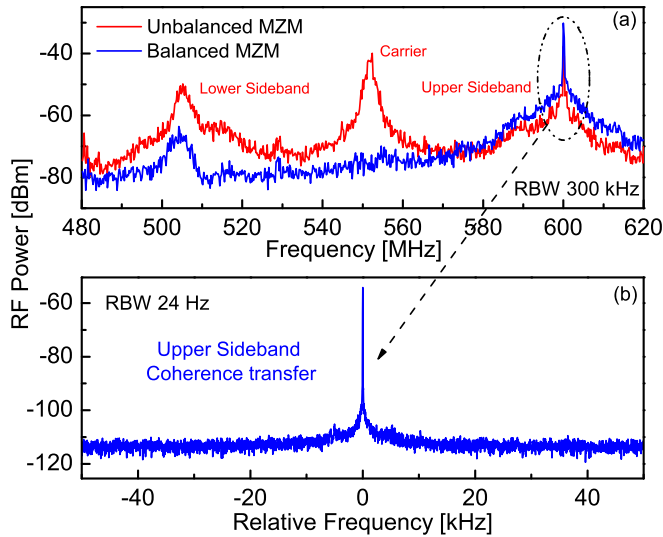}}}%}
\caption{\label{fig:beating} Beating of the MZM output with the VCOF master laser. Upper panel: Broad band view showing the carrier and the other sideband which are strongly suppressed after optimization of the MZM control DC bias levels. Lower panel: High resolution spectrum of the downshifted beat note, still limited by the instrumental resolution (24\,Hz). 58\% of the power is calculated to be in the carrier, corresponding to 0.75\,rad r.m.s. phase noise.
}
\end{figure}

As a first test we use a DFB diode tunable in the proximity of the master laser (1617\,nm) allowing to direct beating of the MZM output against the master. The result in the upper panel of Fig.\ref{fig:beating}, shows an instrument--limited 300\,kHz peak width for the upper sideband and a 4\,MHz-wide peak for the lower sideband, i.e. twice as large as the carrier peak. In order to make this last visible, the MZM must be purposely unbalanced, which allows to change the relative peak intensities while preserving their widths.

To quantify the width of the coherent peak with better resolution we downshift the beating note via a frequency mixer in order to sample the signal using a 200\,MHz GAGE acquisition card (equipped with a 100\,MHz low pass filter). By Fourier transform this produces the higher resolution spectrum in the bottom panel of Fig.\ref{fig:beating}, a beat note with Fourier--transform--limited 24\,Hz width, lying 57\,dB above a flat pedestal. By numerical integration over a span of 60\,MHz, we estimate that about 58\% of the power is in the carrier, corresponding to 0.75\,rad r.m.s. phase noise.

\begin{figure}
\centerline{\rotatebox{270}{\scalebox{0.36}{\includegraphics{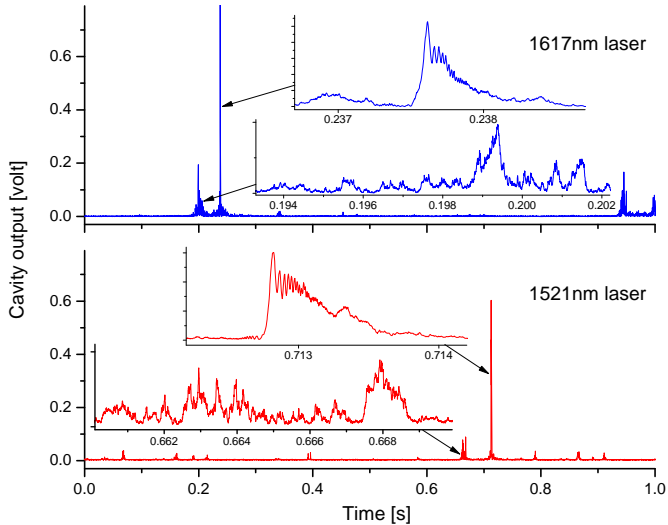}}}}
\caption{\label{fig:cavity_output} A 450\,000 finesse cavity used as an optical spectrum analyzer. 1\,s on the horizontal scale is one cavity FSR (435\,MHz). Top: coherence transfer applied to a DFB laser emitting close to the master VCOF laser. In the insets are detailed views of the narrow coherent peak from the upper sideband and the broad peak from the lower sideband. Bottom: Same observations when using a DFB laser lying at the far edge of the comb.}
\end{figure}

In order to test the coherence transfer over a broad spectral range we use as an optical spectrum analyzer a high finesse optical cavity whose length is linearly scanned by a piezoelectric actuator. This is 34.5\,cm long (Free Spectral Range FSR=435\,MHz) and has finesse 450\,000 around 1610\,nm, corresponding to about 1\,kHz wide resonances. We add a fibered optical amplifier at the MZM output to increase the cavity injection signal, followed by an optical isolator before mode--matched cavity injection. The cavity length is slowly (1\,Hz) modulated over one FSR delivering the transmission transients of the top panel of Fig.\ref{fig:cavity_output} with the same slave laser as before (1617\,nm). The strong narrow peak corresponds to the upper sideband, while the weaker broad peak is associated with the other sideband.
Smaller peaks are due to residual transverse modes (narrow or broad, depending on the sideband they correspond to). By zooming on the narrow sideband transient, a well--known chirped ringing pattern is visible, which can occur only if the incident field is narrower than the cavity mode\cite{Morville2002b}. This is produced by the intracavity field buildup at the passage through resonance, which then exponentially decays and beats with the incident field that continues to weakly feed the cavity off-resonance. The chirp is due to the frequency scan of the cavity resonance. The other sideband does not show any ringing, as is observed when injecting the uncorrected DFB laser directly into the cavity.
The presence of a ringing pattern over a noisy pedestal demonstrates that a fraction of the MZM output power fits inside a sub--kHz bandwidth, consistently with the above result from direct lasers beating (Fig.\ref{fig:beating}).

When changing DFB laser, the same cavity output patterns are observed up to the low wavelength edge of our comb. In particular, the bottom panel of Fig.\ref{fig:cavity_output} shows signals from a 1521\,nm DFB laser, which are equivalent to those from the 1617\,nm laser. Therefore, at the kHz level of the scanning cavity mode width, the coherence transfer appears to be as good for a laser at the VCOF frequency (186\,THz) as for a laser at the comb edge (198\,THz). We should underline that over this range the reflectivity of the cavity mirrors is almost constant, as is deduced from the exponential envelope of the coherent peaks in Fig.\ref{fig:cavity_output} which display a decay constant of about 150$\,\mu$s for both frequencies. In other words, if a degradation of the coherence transfer is present, it stays below 1\,kHz over the explored 12\,THz range.

This is consistent with the frequency noise analysis of Erbium combs previously reported\cite{Newbury2007}, where the noise around the optical carrier is shown to be dominated by environmentally--induced laser cavity length fluctuations. According to the elastic tape picture of the comb noise\cite{Telle2002}, such fluctuations force the comb modes to breath around a fixed point at nearly zero frequency, making their linewidth to increase by a factor of about 0.05\,kHz per THz (at 1\,ms observation time, longer than the photon decay time of our high--finesse cavity). Thus, within a 12\,THz frequency span the non--common--mode frequency noise contribution of the comb can be estimated to about 0.6\,kHz.

In conclusion, we demonstrate a simple and robust FF approach for phase locking two CW single mode lasers beating with the same OFC, with a fast sideband modulator as the only optical control element.
As we demonstrate, this can be applied to high precision cavity--enhanced spectroscopy over the 12\,THz spectral range of an Erbium fiber comb. A set of cost--effective widely tunable telecom lasers can thus be used to interrogate molecular absorption in a high finesse cavity without any trade off in terms of precision, sensitivity and spectral resolution, these being inherited from a single high coherence master (e.g. our VCOF system). In our experimental conditions the coherence transfer was hampered by a comb mode spacing of 100\,MHz: This forced the comb--DFB beat note to be singled out from the replicas due to the adjacent comb modes by means of a relatively tight bandpass filter, slicing off a portion of the DFB noise spectrum. Using a comb with wider mode spacing, effective control bandwidths exceeding 100\,MHz are within reach as recently demonstrated by direct FF lock of a DFB laser against a VCOF laser\cite{Bukart2014a}. 
While we used a commercial comb with standard low--bandwidth controls of $f_0$ and $f_{rep}$, for demanding applications where the contribution from the $f_{rep}$ noise is inacceptable, only this comb parameter would need to be stabilized by a high--bandwidth servo control system.
%It should be noted that the power loss associated with phase--modulated sideband generation could be compensated by an optical amplifier without penalty for the coherence transfer if this is inserted before the splitter feeding the modulator and the comb beating line.
Even though at present the technique works in the telecom range where fibered MZM modulators are readily available, it will eventually become exploitable in the visible or mid--infrared regions thanks to progress with guided optics devices based on lithium--niobate or silicon technologies, respectively.

The authors recognize financial support by: Polo di Lecco - Politecnico di Milano, Institute of Photonics and Nanotechnology of CNR, LabexOSUG@2020 project (ANR10 LABX56), P\^ole SMINGUE (Universit\'e Joseph Fourier), Femto network of CNRS.

% Bibliography
%\bibliographystyle{osajnl} %ol
%\bibliography{bib}

\end{document}